\documentstyle[12pt]{article}
\begin{document}
\newcommand{\reals}{\mbox{${\rm I\!R }$}}
\newcommand{\komplex}{\mbox{${\rm I\!\!\!C }$}}
\newcommand{\nats}{\mbox{${\rm I\!N }$}}
\newcommand{\intgs}{\mbox{${\rm Z\!\!Z }$}}
\newcommand{\cab}{{\cal B}}
\newcommand{\can}{{\cal N}}
\newcommand{\cam}{{\cal M}}
\newcommand{\caz}{{\cal Z}}
\newcommand{\cao}{{\cal O}}
\newcommand{\cac}{{\cal C}}
\newcommand{\cah}{{\cal H}}
\newcommand{\al}{\alpha}
\newcommand{\be}{\beta}
\newcommand{\de}{\delta}
\newcommand{\ep}{\epsilon}
\newcommand{\ga}{\gamma}
\newcommand{\la}{\lambda}
\newcommand{\om}{\omega}
\newcommand{\ze}{\zeta}
\newcommand{\De}{\Delta}
\newcommand{\Om}{\Omega}
\newcommand{\Si}{\Sigma}
\newcommand{\rS}{{\rm S}}
\newcommand{\snnu}{\sum_{n=0}^{\infty}}
\newcommand{\slnu}{\sum_{l=0}^{\infty}}
\newcommand{\snuu}{\sum_{n=-\infty}^{\infty}}
\newcommand{\sluu}{\sum_{l=-\infty}^{\infty}}
\newcommand{\sied}{\sum_{i=1}^3}
\newcommand{\sneu}{\sum_{n=1}^{\infty}}
\newcommand{\pxi}{\prod_{i=1}^3\left(1-e^{-x_i n}\right)}
\newcommand{\pxj}{\prod_{j=1}^3\left(1-e^{-x_ j n }\right)}
\newcommand{\en}{e^{-n\ep\al}}
\newcommand{\enx}{e^{-n\ep\al-nx_i}}
\newcommand{\exi}{(1-e^{-nx_i})}
\newcommand{\exj}{(1-e^{-nx_j})}
\newcommand{\xij}{x_1x_2+x_1x_3+x_2x_3}
\newcommand{\xp}{x_1x_2x_3}
\newcommand{\zrz}{\zeta_R (2)}
\newcommand{\zrd}{\zeta_R (3)}
\newcommand{\zrv}{\zeta_R (4)}
\newcommand{\ent}{d (\nu )}
\newcommand{\pnenj}{\prod_{j=1}^p\nu_j}
\newcommand{\nn}{\nonumber}
\renewcommand{\theequation}{\mbox{\arabic{equation}}}
\newcommand{\intsi}{\int\limits_{\Sigma}d\sigma_x\,\,}
\newcommand{\back}{\bar{\Phi}}
\newcommand{\coba}{\bar{\Phi}^{\dagger}}
\newcommand{\abl}{\partial}
\newcommand{\pa}{\partial}
\newcommand{\qpi}{(4\pi)^{\frac{q+1} 2}}
\newcommand{\snenp}{\sum_{n_1,...,n_p=0}^{\infty}}
\newcommand{\tint}{\int\limits_0^{\infty}dt\,\,}
\def\beq{\begin{eqnarray}}
\def\eeq{\end{eqnarray}}
\newcommand{\zb}{\zeta_{{\cal B}}(}
\newcommand{\rzb}{Res\,\,\zb}
\newcommand{\zn}{\zeta_{{\cal N}}\left(} 
\newcommand{\rzn}{Res\,\,\zn}
\newcommand{\fr}{\frac}
\newcommand{\sip}{\frac{\sin (\pi s)}{\pi}}
\newcommand{\rzs}{R^{2s}}
\newcommand{\g}{\Gamma\left(}
\newcommand{\ikma}{\int\limits_{\ga}\frac{dk}{2\pi i}k^{-2s}\frac{\pa}{\pa k}}
\newcommand{\suani}{\sum_{a=0}^i}
\newcommand{\zem}{\zeta_{{\cal M}}}
\newcommand{\hem}{A^{\cam}}
\newcommand{\hen}{A^{\can}}
\newcommand{\man}{{\cal M}}
\newcommand{\pold}{D^{(d-1)}}
\newcommand{\zesd}{\zeta_{S^d}}
\newcommand{\fac}{\frac{(4\pi)^{D/2}}{a^d|S^d|}}
\newcommand{\sri}{\sum_{i=1}^ d  r_i}
\newcommand{\pri}{\prod_{i=1}^d r_i}
\newcommand{\ber}{B^{(d)}}
\newcommand{\ar}{a|\vec r )}
\newcommand{\Ga}{\Gamma^a}
\newcommand{\Gb}{\Gamma^b}
\newcommand{\sumtn}{\sum_{\vec n_t \in \intgs^{d-1}/\{0\}}}
\newcommand{\sumnd}{{\sum_{n_d=-\infty}^\infty}\!\!\!^\prime} 
\newcommand{\sumt}{\sum_{\vec n\in \intgs^{d}/\{\vec 0\}}}
\newcommand{\ajabc}{A_j^{a,b,c}}
\newcommand{\res}{\mbox{Res }}
\newcommand{\ug}{u+gn_d}
\newcommand{\smp}{\sum_{m=0}^p \sum_{l=0}^m \gamma_{ml}^p }
\newcommand{\smpe}{\sum_{m=0}^{p-1}\sum_{l=0}^m \de_{ml}^p }
\newcommand{\smpen}{\sum_{m=0}^{p-1}\sum_{l=0}^m \de_{ml0}^p }
\newcommand{\smpee}{\sum_{m=0}^{p-1}\sum_{l=0}^m \de_{ml1}^p }
\newcommand{\G}{\Gamma}
\newcommand{\ph}{\phantom}
\newcommand{\nt}{\vec n _t^2}
\begin{titlepage}

\title{The $a_{3/2}$ heat kernel coefficient for oblique boundary conditions} 
\author{J.S. Dowker\thanks{Electronic address: dowker@a3.ph.man.ac.uk} 
\phantom{a}and 
Klaus Kirsten\thanks{Electronic address: klaus.kirsten@itp.uni-leipzig.de}\\
\\
$^*$ Department of Theoretical Physics,\\
The University of Manchester, Manchester, England\\
\\
$^\dagger$ Universit\"{a}t Leipzig, 
Institut f\"{u}r Theoretische Physik,\\
Augustusplatz 10, 04109 Leipzig, Germany}

\maketitle

\begin{abstract}
We present a method for the 
calculation of the $a_{3/2}$ heat kernel 
coefficient 
of the heat operator trace for a partial differential operator of 
Laplace type on a compact Riemannian manifold with 
oblique boundary conditions. Using special case
evaluations, restrictions are put on the general form of the coefficients,
which, supplemented by conformal transformation techniques, allows the 
entire smeared coefficient to be determined.
\end{abstract}
\end{titlepage}

\section{Introduction}
The general topic of heat-kernel expansions and eigenvalue asymptotics 
of an operator $L$ on a $D$-dimensional Riemannian manifold $\cam$ has been 
an important issue for more than $20$ years. In mathematics this 
interest stems, in particular, from the well-known connection that 
exists between the heat-equation and the Atiyah-Singer index theorem 
\cite{1}. In physics the expansion stands out in
different domains of quantum field theory, as it contains, for example,
information on the scaling and divergence behaviour, \cite{2,3,4,5,6}.

Of particular interest is the case where $L$ is a Laplacian-like 
operator. When the manifold $\cam$ has a boundary $\partial \cam$ 
one has to impose boundary conditions that guarantee the 
self-adjointness and the ellipticity of the operator $L$. The 
traditional, and simplest, conditions are Dirichlet and Neumann, 
and also a generalization of the latter referred to as Robin conditions. 
In this case, the heat-kernel coefficients $a_k$ have both a volume and 
a boundary part \cite{7,8}. Writing the expansion in its usual form
\beq
K(t) \sim \sum_{k=0,1/2,1,...}\quad a_k t^{k-D/2} \label{I1}
\eeq
one has the split
\beq
a_k = \int_{\cam} dx\,\, b_k + \int_{\partial \cam}dy \,\,c_k.\label{I2}
\eeq 
Here, the volume part, $b_k$, does not depend on the boundary conditions, 
whereas $c_k$ exhibits a nontrivial dependence on the chosen boundary 
condition. 

Whereas the calculation of the volume part nowadays is nearly 
automatic \cite{9,10,11}, the analysis of $c_k$ is in general much more 
difficult. Only quite recently has the coefficient $c_2$ for Dirichlet 
and for Robin boundary conditions been found \cite{12,13,14,15,16,17,18}.
$c_{5/2}$ for manifolds with totally geodesic boundaries is given in \cite{19}.

One of the approaches is based on the functorial methods most 
systematically used by Branson and Gilkey \cite{12}. Conformal transformation 
techniques give relations between the numerical multipliers in the 
heat-kernel coefficients. However, on its own, this method is unable to 
determine the coefficients fully. Additional information is needed, 
coming from other functorial relations or special case calculations \cite{12}.

In contrast to this traditional boundary conditions relatively little is known 
about{\it oblique boundary conditions}. These more general conditions
takes the form
\beq
{\cal B}=\nabla_N+{1\over2}\big(\Ga\widehat\nabla_a+\widehat\nabla_a
\Ga\big)-S, \label{I3}
\eeq
and involves tangential (covariant) derivatives, $\widehat\nabla_a$, 
computed from the induced metric on the boundary, 
$\Ga$ is a bundle endomorphism valued boundary vector field and 
$S$ is a hermitian bundle automorphism. Finally, 
$\nabla_N$ is the {\it outward} normal derivative at the boundary.
In order to ensure symmetry of the operator $L$ together with the boundary 
condition 
\beq
{\cal B}V\big|_{\pa\man}=0, \label{I4}
\eeq
on a section of some vector bundle, one has to impose 
$(\Gamma^a)^\dagger = -\Gamma ^a$ and $S^\dagger = S$. This kind of boundary 
conditions arises naturally if one requires invariance of the boundary 
conditions under infinitesimal diffeomorphisms \cite{20,21,22} or 
Becchi-Rouet-Stora-Tyutin transformations \cite{23}. Furthermore 
(\ref{I4}) is suggested by self-adjointness theory \cite{24,25} and string
theory \cite{26,27}.

Although these boundary conditions have been subject of classical analysis 
(see e.g. \cite{1,28,29,30}) the explicit determination of the heat-kernel
coefficients has hardly begun. Some of the lower coefficients 
(up to $a_1$) have been
evaluated by McAvity and Osborn \cite{24} using their extension of
the recursion method developed by De Witt for closed
Riemannian manifolds. Another
approach has been expounded in \cite{31} based on the functorial
methods used in \cite{12}. In \cite{31} the coefficients in the very special
case of a flat ambient manifold with a totally geodesic, flat boundary 
have been computed. However, the combined knowledge of the coefficient 
with a totally geodesic, flat boundary and the functorial method relations 
does not allow for the determination of the full (under certain
assumptions, see section II) $a_{3/2}$ coefficient. The reason behind this
is that no knowledge of the extrinsic curvature terms at the boundary
is obtained via this example. This hinders the determination of all the 
universal constants.

It is here that the approach of special case evaluation on manifolds 
with non-vanishing extrinsic curvature becomes important. A specific class
of manifolds with this property is the generalized cone, a particular 
curved manifold whose boundary is not geodesically embedded. For this 
kind of manifold techniques have been developed earlier for the evaluation
of heat-kernel coefficients and functional determinants \cite{32}
(see also \cite{33,34,35,36,37}) and have been applied to Dirichlet and 
Robin boundary conditions. Recently the generalisation to smeared 
heat-kernel coefficients was found \cite{38}.

This generalisation turns out to be important because functorial 
techniques (apart from other things) yield relations between the 
smeared and non-smeared case. The information one can get on the 
``smeared side" (these are terms containing normal derivatives of the 
smearing function $f$) is crucial to find the full ``non-smeared" 
side (even present for $f=1$). This has been demonstrated by determining 
the full $a_{5/2}$ coefficient containing the whole group of extrinsic 
curvature terms \cite{39,40}.

When employing the formalism of \cite{32,33} to oblique boundary 
conditions, the tangential derivatives in (\ref{I3}) cause added 
complication. For the case of the $4$ dimensional ball these have been
overcome in \cite{39} and the coefficients up to $a_2$ were found for 
constant $\Gamma^i$ (for new considerations on this specific manifold
see \cite{dimaeli}). Here we will continue this analysis by generalizing
it to the arbitrary dimensional as well as to the smeared case. As a 
result enough universal constants will emerge from special cases in order 
to find the full $a_{3/2}$ coefficient for oblique boundary conditions. 

The organisation of the paper is as follows. In the next section we     
state in detail under which assumptions we are going to 
determine the $a_{3/2}$ coefficient (purely Abelian problem, covariantly 
constant $\Gamma^i$). For this situation the general form of the coefficients
has been stated in \cite{31} and we explain which universal constants and
which relations among them can be obtained from the generalized cone.
In section 3 the explicit calculation on the cone is performed and the 
information predicted found. Having this information at hand the functorial
techniques \cite{12} are applied to this boundary condition \cite{31}.
To obtain as much information as possible even for the case of 
covariantly constant $\Gamma^i$ the structure and conformal properties 
of the non-covariantly constant $\Gamma^i$ will be needed and displayed.
At the end, we will see that one relation is missing which will be 
obtained in section 5 by dealing with the manifold $B^2\times T^{d-1}$
($B^2$ being the two dimensional ball and $T^{d-1}$ the 
$(d-1)$-dimensional torus).  The conclusions summarize the 
main ideas and give further possible applications. 

\section{Restrictions from the generalized cone}
Before actually doing the calculation on the generalized cone we discuss
the general structure of the coefficients \cite{31} and see 
what restrictions can be obtained from the coefficients found on our 
specific manifold. First one has to state clearly the assumptions under
which the structure of the coefficients is formulated. We follow here 
\cite{31}, otherwise the situation is considerably 
more complicated \cite{41}. The assumptions are as follows:

(i) The problem is purely Abelian, i.e. the matrices $\Gamma^i$ commute  
: $[\Gamma^i, \Gamma^j ] =0$.

(ii) The matrix $\Gamma^2 = h_{ij} \Gamma^i \Gamma^j$ which automatically 
commutes with $\Gamma^j$ by virtue of (i), commutes also with the 
matrix $S$: $[\Gamma ^2, S ] =0$.

(iii) The matrices $\Gamma^i$ are covariantly constant with respect to the 
(induced) connection on the boundary: $\widehat\nabla_i \Gamma^j =0$.

Under these assumptions we consider the Laplace-like 
operator 
\beq
L=-g^{ij} \nabla_i \nabla_j - E \label{II1}
\eeq
where $E$ is an endomorphism of the smooth vector bundle $V$ over $\cam$ 
and $\nabla$ is a connection, together with the boundary conditions (\ref{I3}).
Then the general form of the heat-kernel coefficients is 
\beq
a_{1/2} (f) &=& (4\pi )^{-(D-1)/2} Tr(\delta f) [\pa \cam ] \label{II2}\\
a_1 (f) &=& (4 \pi )^{-D/2} \frac 1 6 \left\{ 
               Tr (6fE+fR) [\cam] \right.\label{II3}\\
& &\left.+Tr \left\{f(b_0 K +b_2 S) +b_1 f_{;N}+
         f\sigma_1 K_{ab} \Ga \Gb\right\}
         [\pa\cam]\right\} \nn\\
a_{3/2} (f) &=& (4\pi )^{-(D-1)/2} \frac 1 {384} Tr \left[f(c_0 E+c_1 R+c_2 
R^a_{\phantom{a}NaN}+c_3 K^2+c_4 K_{ab}K^{ab} \right.\nn \\
& &\left. +c_7 SK +c_8 S^2)
 +f_{;N} (c_5 K +c_9 S) +c_6 f_{;NN} \right] [\pa\cam] \label{II4}\\
& &+Tr\left[ f (\sigma_2 (K_{ab} \Ga \Gb ) ^2 +\sigma_3 K_{ab} \Ga \Gb K 
+\sigma_4 K_{ac} K^c_b \Ga \Gb  \right.\nn  \\
& & +\lambda_1 K_{ab} \Ga \Gb S 
 + \mu_1 R_{aNbN} \Ga \Gb +\mu_2 R^c_{\phantom{c} acb} \Ga \Gb  \nn\\
& &\left.+b_1 \Omega_{aN} \Ga ) 
+\beta_1 f_{;N} K_{ab} \Ga \Gb \right] [\pa\cam] \nn
\eeq
Here and in the following $f[\cam]=\int_{\cam}dx\, f(x) $
and $f[\partial \cam] = \int_{\partial \cam} dy f(y) $,
with $dx$ and $dy$ being the Riemannian volume elements
of $\cam$ and $\partial \cam$. In addition, the semi-colon denotes 
differentiation
with respect to the Levi-Civita connection of $\cam$, $\Omega$ is the 
connection of $\nabla$ and $K_{ab}$ the extrinsic curvature. 
Finally, our sign convention for the Riemann tensor is 
$R^i_{\phantom{i}jkl} = -\Gamma^i_{jk,l}+
\Gamma^i_{jl,k} +
\Gamma^i_{nk} \Gamma^n_{jl} -
\Gamma^i_{nl} \Gamma^n_{jk}
$ (see for example \cite{42}).

Although $a_{1/2}$ and $a_1$ have been determined previously \cite{24,31,39},
we have included them in the list to explain clearly our procedure.
The terms in $a_{3/2}$ are grouped together such that the first two lines,
$c_0$ up to $c_9$, contain the type of geometric invariants already present 
for Robin boundary conditions, whereas all the other terms are due only to
the tangential derivatives in the boundary condition.

Our aim in the next section will be to put restrictions on the universal 
constants of eqs. (\ref{II2})-(\ref{II4}) by calculating the coefficients 
of the conformal Laplacian ($E=-(d-1)R/(4d)$)
on the bounded generalized cone. By this we mean the $D=(d+1)$-dimensional 
space $\cam = I \times \can$ with the hyperspherical metric \cite{43}
\beq
ds^2= dr^2+r^2 d\Sigma^2 , \label{II5}
\eeq
where $d\Sigma^2$ is the metric on the manifold $\can$ and $r$ runs from 
$0$ to $1$. $\can$ will be referred to as the base of the cone. If it has 
no boundary then it is the boundary of $\cam$.

It is clear that a special case calculation will be simplified considerably 
by taking a constant $\Ga$, say $\Gamma^d = ig$, with the real constant $g$. 
In order that this is covariantly constant one might think of taking a flat 
base manifold $\can$. The most natural such manifold where much is known 
about all
required spectral propoerties is the torus. We thus choose $\can = 
T^d$, namely the equilateral $d$-dimensional torus with perimeter $L=2\pi$
and metric $d\Sigma^2 = dx_1^2+...+dx_d^2$. 
Its volume is ${\rm vol} (T^d) = (2\pi )^d$ and the basic geometrical tensors
read 
\beq
R^{ij}_{\phantom{ij}kl} = \frac 1 {r^2} (\delta^i_l \delta ^j_k 
                           -\delta^i_k \delta^j_l ), \quad 
          K^a_b = \delta^a_b   .\nn
\eeq
Furthermore, we choose a specific smearing function $f=f(r)$ which will 
allow the 
calculation to be effected but which contains nevertheless all the information 
concerning the universal constants one can obtain. A possible choice is 
\cite{38} (see section 4 for the applications in the given context)
\beq
f(r) = f_0 +f_1 r^2+f_2 r^4.\nn
\eeq
For this special setting, the coefficients will have the following appearance:
\beq
\frac{(4\pi)^{d/2} }{(2\pi)^d} a_{1/2} (f)&=&\delta f(1) \label{II6}\\
\frac{(4\pi)^{D/2} }{(2\pi)^d}6 a_1 (f) &=&\frac 1 2 (d-3)(d-1) \left[
    \frac{f_0}{d-1} +\frac{f_1} {d+1} +\frac{f_2}{d+3}\right] \label{II7}\\
        & &+b_0 f(1) d +b_1 f_{;N} (1) +b_2 f(1) S -\sigma_1 f(1) g^2 \nn\\
\frac{(4\pi)^{d/2} }{(2\pi)^d} 384a_{3/2} (f)&=& 
             f(1) \left[
       c_0 (d-1)^2/4 -c_1 d(d-1) +c_3 d^2 +c_4 d +c_7 S d\right. \nn\\
      & &+c_8 S^2 +\sigma_2 g^4 -\sigma_3 d g^2 -\sigma_4 g^2 \nn\\
      & & \left.-\lambda_1 S g^2 -\mu_2 g^2 (1-d) \right] \label{II8}\\
     & & +f_{;N} (1) \left[c_5 d +c_9 S-\beta_1 g^2\right]
                 +c_6 f_{;NN}(1)  \nn
\eeq
Thus, by doing the calculation on the manifold $\cam = I\times T^d$ with 
$f(r) = f_0 +f_1 r^2+f_2 r^4$ and by comparing terms containing a specific 
number of normal derivatives of $f$ together with a fixed number of 
powers in $d$ and $S$ the following information can be extracted,
\beq
a_{1/2} & & \delta,\nn\\
a_1 & & b_0,b_1,b_2,\sigma_1 ,\nn\\
a_{3/2} & & c_3-c_1+c_0/4, \Gamma^2 (\sigma_3-\mu_2 ) +c_4+c_1-c_0/2, 
          c_5,c_6,c_7,c_8,c_9,
             \beta_1,\lambda_1, \nn\\
         & &\Gamma^4 \sigma_2 +\Gamma^2 \sigma_4 +\Gamma^2 \mu_2
              +c_0 /4 . \nn
\eeq
The considerable amount of information that one derives from this example 
is apparent. $a_{1/2}$ and $a_1$ are completely determined without any 
additional input and from $a_{3/2}$ one gets $10$ of $18$ unknowns.
There is good hope that the remaining information can be found from  
functorial techniques. 

Having, therefore, a good motivation, we can embark on special case 
calculations and see afterwards if the functorial 
relations can complete the information.

\section{Non-smeared generalised cone calculation}
So let us turn to the spectral analysis of the conformal
Laplacian on the described 
generalized cone together with the boundary condition (\ref{I3}). The 
conformal Laplacian is 
\beq
\Delta_{\cam} -\frac{d-1}{4d} R = 
  \frac{\pa ^2}{\pa r^2} +\frac d r \frac{\pa}{\pa r} 
   +\frac{(d-1)^2}{4r^2}+\frac 1 {r ^2} \Delta_{\can} \label{III1}
\eeq
with eigenfunctions
\beq
\frac{J_{\nu} (\alpha r)}{r^{(d-1)/2} } \exp\{i(x_1n_1+...+x_dn_d)\}, \quad
\vec n\in\intgs^d .\label{III2}
\eeq
The index $\nu$ equals 
\beq
\nu = \left( n_1^2 +...+n_d^2  \right) ^{1/2} 
\label{III3}
\eeq
and the eigenvalues $\alpha$ are determined through 
(\ref{I3}) by 
\beq
\alpha J_{\nu} ' (\alpha) +(u+gn_d) J_{\nu}(\alpha) =0. \label{III4}
\eeq
Here $u=1-D/2-S$.

For the determination of the heat-kernel coefficients we follow the approach 
developed in \cite{32,33}. The basic object is the zeta function of $\cam$,
\beq
\zeta (s) = \sum \alpha^{-2s} \label{III5}
\eeq
and the relation 
\beq
a_{k/2} = \Gamma ((D-k)/2) {\rm Res} \zeta_{\cam} ((D-k)/2 ) \label{III6}
\eeq
between the coefficients and the zeta function is used. In addition the 
Epstein type zeta function defined by 
\beq
E_k (s) = \sum_{\vec n \in \intgs^d/\{\vec 0\}} 
        \frac{n_d^k}{\left(n_1^2 +...+n_d^2 
                       \right)^{s} } \label{III7}
\eeq
will turn out to be very useful. Obviously it is connected with the 
spectrum of the Laplacian on the base manifold $\can$, the  
$n_d$-powers arise from the tangential derivatives in (\ref{I3}). 

The starting point of the analysis of $\zeta_{\cam}$ is the  contour integral 
representation,
\beq
\zeta (s) = \sum_{\vec n \in \intgs^d} 
\int_{\gamma} \frac{dk}{2\pi i} k^{-2s} 
\frac{\pa}{\pa k} \ln \left( kJ_{\nu} ' (k) + (u+gn_d) J_{\nu} (k) \right),
\label{III8}
\eeq
where $\gamma$ must enclose all the solutions of (\ref{III4}) on 
the positive real axis. It is the appearance of the $n_d$ in the last term 
that causes the added complications compared to \cite{32,33}.

In the following analysis the index $\nu = \vec 0$ would require a 
separate treatment. Its contribution has the rightmost pole at $s=1/2$ 
because it is associated with the zeta function of a second order 
differential operator in one dimension. Because we are dealing with arbitrary
dimensions this pole (and all other poles to the left of it) are 
irrelevant for our goal due to the relation (\ref{III6}). For convenience 
therefore, we will continue  without including the $\nu = \vec 0$ 
contribution and will still use the same notation, $\zeta (s)$.

Shifting the countour to the imaginary axis, the zeta function 
(with the zero mode $\nu = \vec 0$ omitted, as explained) reads 
\beq
\zeta (s) = {\sin \pi s\over \pi}\sumt\int_0^{\infty} dz \,\, 
(z\nu)^{-2s} {\partial \over \partial z}
\log z^{-\nu}\bigg[ z\nu I_{\nu} ' (z\nu ) + 
(u+gn_d) I_{\nu} (z\nu ) \bigg]. \,\,\label{III9}
\eeq
As shown in detail in \cite{32,33}, the heat-kernel coefficients are
determined solely by the asymptotic contributions of the Bessel functions
as $\nu\to\infty$. In the given consideration more care is
needed since terms like $n_d/\nu$ have to be counted as of order $\nu^0$.

Using the uniform asymptotic expansion of the Bessel function \cite{44} one
encounters the expression 
\beq
\ln\bigg\{1+\bigg(1+\frac{gn_d}{\nu} t \bigg)^{-1}
\bigg[\sum_{k=1}^{\infty} \frac{v_k (t) }{\nu^k}  +\frac{ut}{\nu}
+\bigg( \frac{u+gn_d}{\nu}\bigg)t\sum_{k=1}^{\infty} \frac{u_k(t)} {\nu^k}
\bigg]\bigg\} &=&
\nn\\
& &\hspace{-3cm}\sum_{j=1}^{\infty} {T_j (u,g,t)\over \nu^j}
\label{III10}
\eeq
whereby the $T_j$ are defined and $t=1/\sqrt{1+z^2}$. For the Olver
polynomials, $u_k$ and $v_k$, see \cite{44}.

Asymptotically one finds
\beq
\zeta (s) = A_{-1} (s) +A_0 (s) +A_+ (s) +\sum_{j=1}^{\infty} A_j (s),
\label{III11}
\eeq
where $A_{-1} (s)$ and $A_0 (s)$ are the same as in Robin
boundary conditions \cite{32}, namely
\beq
A_{-1} (s) &=& \frac 1 {4\sqrt{\pi}} \frac{\Gamma (s-1/2)} {\Gamma (s+1) }
E_0 (s-1/2) , \label{III12}\\
A_0 (s) &=& \frac 1 4 E_0 (s) .\label{III13}
\eeq
The new quantities are
\beq
A_+ (s) = {\sin \pi s\over \pi} \sumt\int_0^{\infty} dz \,\, 
(z\nu)^{-2s} {\partial \over \partial z}
\ln\left( 1+{gn_dt\over\nu} \right) ,
\label{III14}
\eeq
and 
\beq
A_j (s)  = {\sin \pi s\over \pi} \sumt\int_0^{\infty} dz \,\, 
(z\nu)^{-2s} {\partial \over \partial z}
{T_j (u,g,t)\over \nu^j} .
\label{III15}
\eeq
In order to proceed it is convenient to express $T_j$ as the finite sum
\beq
T_j = \sum_{a,b,c} f_{a,b,c}^{(j)} {
 \de^c t^a \over \left( 1+\de t\right)^b },
\label{III16}
\eeq
with $\delta = gn_d/\nu$.
The $f_{a,b,c}^{(j)} $ are easily determined via an algebraic
computer programme.

The next steps are to perform the $z$-integrations by the identity,
\beq
\int_0^{\infty} dz \,\, z^{-2s} {zt^x\over (1+\delta t)^y }
= \frac 1 2 {\Gamma (1-s) \over \Gamma (y) } \sum_{k=0}^{\infty}
(-1)^k {\Gamma (y+k) \Gamma (s-1+(x+k)/2) \over k! \Gamma ((x+k)/2) }
\delta ^k ,
\label{III17}
\eeq
and then do the $\vec n$-summation to write everything in terms of the 
Epstein functions (\ref{III7}). Performing these steps one gets first 
\beq
A_+ (s) &=& \frac 1 {2\Gamma (s)} \sum_{n=1}^{\infty} 
\frac{\Gamma (s+n) }{\Gamma (n+1)} E_{2n} (s+n) g^{2n}. \label{III18}
\eeq
For $c$ even and $c$ odd in (\ref{III16}) slightly different representations 
appear such that some more notation is unfortunately necessary. We write
\beq
A_j (s) = \sum_{a,b,c} f_{a,b,c}^{(j)} A_j^{a,b,c} (s) \label{III19}
\eeq
and get
\beq
A_j^{a,b,c} (s) &=& -\frac 1 {\Gamma (s) }  
\sum_{n=0}^{\infty} \frac{\Gamma (b+2n)}{\Gamma (b) \Gamma (2n+1)}
\frac{\Gamma (s+a/2+n)}{\Gamma (a/2 +n) } \nn\\
& &\hspace{2cm} E_{2n+c} (s+n+(j+c)/2) g^{2n+c}.
\label{III20}
\eeq
for $c$ even and 
\beq
A_j^{a,b,c} (s) &=&\frac 1 {\Gamma (s) }    
\sum_{n=0}^{\infty} \frac{\Gamma (b+2n+1)}{\Gamma (b) (2n+1)!}
\frac{\Gamma (s+(a+1)/2+n)}{\Gamma ( (a+1)/2 +n) } \nn\\
& &\hspace{2cm} E_{2n+c+1} (s+n+(j+c+1)/2) g^{2n+c+1}.
\label{III21}
\eeq
for $c$ odd.

We need the residues of $A_{-1},A_0,A_+$ and $A_j$, but this is not too 
difficult, because the Epstein zeta functions are very well studied objects.
For us the relevant properties are 
\beq
E_n (s) =0 \quad \mbox{for }n\mbox{ odd} \label{III22}
\eeq
and for $s=l+d/2$ the residue is 
\beq
\mbox{Res }E_{2l} (l+d/2) = \frac{ \pi^{(d-1)/2} \Gamma (l+1/2)}
{ \Gamma (d/2+l)}  .
\label{III23}
\eeq
A nice feature of the calculation is, that when using the above results 
(\ref{III22}) and (\ref{III23}) in (\ref{III18}) and (\ref{III19}), 
$\mbox{Res }A_+ ((D-k)/2)$ and $\mbox{Res }A_j ((D-k)/2)$ reduce to the 
series representation of the generalized hypergeometric function \cite{45},
\beq
_pF_q (\alpha_1,\alpha_2,...,\alpha_p; \beta_1,\beta_2,...,\beta_q;z) 
\sum_{k=0}^{\infty} \frac{(\alpha_1)_k (\alpha_2)_k ...(\alpha_p)_k}
         {(\beta_1)_k (\beta_2 )_k ... (\beta_q) _k } \frac{z^k}{k!}.
\label{III24}
\eeq
For example, for $A_+$ only one contributions arises which, usefully 
normalized, reads 
\beq
\lefteqn{
\Gamma ((D-1)/2) \frac{(4\pi)^{d/2}}{(2\pi )^d} \mbox{Res }A_+
 ((D-1)/2) } \nn\\
&=& \frac{1}{2}  
\left\{_2F_1 (1/2,d/2; d/2;g^2)-1 \right\}    \nn\\
&=& \frac{1}{2} 
\left\{(1-g^2)^{-1/2} -1\right\} . \label{III25}
\eeq
In this case the intermediate step in terms of the hypergeometric function
is artificial of course but useful in general.

In order to give the contributions of $A_j^{a,b,c}$, we have 
to distinguish between even and odd $j$. Although a bit lengthy we find 
it useful to state these results. Let us stress, that we have already, in 
principle, determined  an arbitrary number of 
heat-kernel coefficients for the Laplacian on a generalized cone with oblique 
boundary boundary conditions in the non-smeared case. For $c$ odd and $j$ 
odd we find
\beq
\lefteqn{
\Gamma ((D-1-j)/2) \frac{(4\pi)^{D/2}}{(2\pi)^d} 
\res\ajabc ((D-1-j)/2) =  }\nn\\
& & 2 
\frac{\Gamma (1+c/2)}{\Gamma ((a+1)/2) \Gamma (b)  } \nn\\ 
& &   \left( \frac{D+c} 2\right)_{\frac{a-j-c} 2}     
g^c \left( \frac d {dg} \right) ^{b-1} g^b \label{III26}\\
& & _3F_2 (1,(d+a+1-j)/2,1+c/2;
           (a+1)/2,(D+c)/2;g^2), \nn
\eeq
which contributes to the coefficient $a_{(1+j)/2} $. For the 
specific values of $b,c,j$ and $k$ needed the hypergeometric function always
reduces to a simple algebraic or a hyperbolic function. The above result 
neatly summarizes all this information in one equation.

For $c$ odd and $j$ even $\ajabc$ also contributes to the coefficient 
$a_{(j+1)/2}$ and the relevant result is $1/(2\sqrt{\pi})$ times the above  
(note that in this case the normalization is 
$(4\pi)^{d/2}$).

Furthermore, for $c$ even and $j$ odd the analogous result is 
\beq
\lefteqn{
\Gamma ((D-1-j)/2) \frac{(4\pi)^{D/2}}{(2\pi)^d}
\res\ajabc ((D-1-j)/2)=} \nn\\
& &- 2  
\frac{\Gamma ((1+c)/2)}{\Gamma (a/2) \Gamma (b)  } \nn\\
& & \left( \frac{d+c} 2\right)_{\frac{a-j-c} 2}     
g^c \left( \frac d {dg} \right) ^{b-1} g^{b-1} \label{III27} \\
& & _3F_2 (1,(d+a-j)/2,(1+c)/2;
           a/2,(d+c)/2;g^2), \nn
\eeq
For $c$ even and $j$ even the same rules as above hold, furthermore the 
same comments. 

The above results allow for a direct evaluation of the coefficients by an
algebraic computer program such as Mathematica. However, before stating the 
results let us describe the necessary modifications when dealing with the 
smeared case.

\section{Generalization to the smeared heat-kernel coefficients}
The inclusion of a smearing function that depends only on the radial variable 
 results in the smeared zeta function,
\beq
\ze(f;s)=\sumt\int_\ga {dk\over2\pi i}k^{-2s}\int_0^1 dr\,f(r)\bar
J^2_\nu(kr) r\,{\pa\over\pa k}\ln (k J_\nu ' (k) + (u+gn_d)J_\nu(k)).
\label{IV1}
\eeq
For $f(r)$ a polynomial, we have shown how to analyse (\ref{IV1}) for 
Dirichlet and Robin boundary conditions in \cite{38} and so for the general 
procedure see this reference. The generalisation to oblique boundary 
conditions is obtained here.

The bar in (\ref{IV1}) signifies normalized. Explicitly 
\beq
\bar J _\nu (kr) = \frac{ \sqrt 2 k}{((u+gn_d)^2+k^2-\nu^2)^{1/2} J_\nu (k) }
J_\nu (kr) .\label{IV2}
\eeq
For 
\beq
f(r) = \sum_{n=0}^N f_n r^{2n} \label{IV3} 
\eeq
we need normalization integrals of the type 
\beq
S[1+2p]=
\int_0^1dr\,\bar J_\nu^2(\al r)r^{1+2p}. \label{IV4}
\eeq
These can be treated using Schafheitlin's reduction formula \cite{46},
which for the present case gives the recursion
\beq
S[1+2p]&=&{2p \over 2p+1} {\nu^2 -p^2  \over \alpha ^2} S[2p-1] \nn\\
   & &+ {1 \over 2p+1} \left( 1+{2p (u+p) \over \alpha^2 +(\ug )^2 -\nu ^2 }
\right), \label{IV5}
\eeq
starting with $S[1]=1$. So $S[1+2p]$ has the following form
\beq
S[1+2p] &=& \sum_{m=0}^p \left({\nu \over \alpha} \right) ^{2m}
\sum_{l=0}^m \gamma _{ml}^p\, \nu ^{-2l} \nn\\
& &+{1 \over \alpha^2 +(\ug)^2 -\nu ^2} \sum_{m=0}^{p-1} \left(
{\nu \over \alpha }\right) ^{2m} \sum_{l=0}^n \delta _{ml} ^p\, \nu^{-2l}.
\label{IV6}
\eeq
The numerical coefficients $\gamma_{ml}^p$ and $\delta_{ml}^p$ are easily 
determined recursively. As a result, apart from characteristic differences,
the smeared zeta function $\ze_\cam (f,s)$ takes a similar form as $\zeta
_\cam (s)$.

It will turn out convenient to devide $\ze_\cam (f,s)$ into different pieces 
characterized below. First, respecting the structure in (\ref{IV6}), we 
define 
\beq
\ze_\ga^p (f,s) &=& \smp \sumt \nu^{2m-2l}\int_\ga \frac{dk}{2\pi i} k 
^{-2(s+m)} \nn\\
& &\frac{\pa}{\pa k} ln (k J_\nu ' (k) + (u+gn_d)J_\nu(k)) \label{IV7}\\
\tilde \ze_\delta^p (f,s) &=& \smpe \sumt \nu^{2m-2l}\int_\ga \frac{dk}{2\pi i}
\frac{k^{-2(s+m)}}{(k^2 +(\ug)^2 -\nu^2)} \nn\\
& &\frac{\pa}{\pa k} ln (k J_\nu ' (k) + (u+gn_d)J_\nu(k)), \label{IV8}
\eeq
where the contour $\gamma$ has to be chosen so as to enclose  the
zeros of {\it only} $ kJ_\nu ' (k) + (\ug ) J_\nu (k) $. 
Thus the poles of $S[1+2p]$, located at
$k=\pm \sqrt{\nu^2 -(\ug)^2}$, must be outside the contour. It is
important to locate the contour properly because, when deforming it to
the imaginary axis, contributions from the 
pole at $k=\sqrt{\nu^2-(\ug)^2}$ arise.

The index $p$ referes to the fact that these are the contributions coming 
from the power $r^{2p}$ in (\ref{IV3}). In order to obtain the full zeta 
function, the $\sum_{p=0}^N f_p \ze^p$ has to be done.

The first piece, $\ze_\ga^p$, may be given just by inspection. Comparing 
(\ref{IV7}) with the non-smeared zeta function (\ref{III8}) the contour integral
is the same as previously once $s\to s+m$ has been put. Due to the additional
factor $\nu^{2m-2l}$ the argument of the base zeta function has to be raised
by $l-m$. For explanatory purposes let us give as an explicit example
\beq
A_{-1}^\ga (f,s) =\frac 1 {4\sqrt{\pi}} 
\sum_{p=0}^N f_p \smp \frac{\Gamma (s+m-1/2)}{\Gamma (s+m+1)} E_0 (s+l-1/2).
\label{IV9}
\eeq
In exactly the same way, $A_0^\ga (f,s)$, $A_+^\ga (f,s) $ and $A_j ^\ga (f,s)$ 
are obtained from (\ref{III13}), (\ref{III18}), (\ref{III20}) and (\ref{III21}).
This is the stage where the properties (\ref{III22}) and (\ref{III23}) are 
used and the contributions to the heat-kernel coefficients in terms of 
hypergeometric function emerge. They will not be displayed, however, 
explicitly, because the structure is the one already seen in (\ref{III26}) 
and the 
way they are obtained is identical to the procedure described in section 3.

We continue with the analysis of $\tilde \ze _\de ^p$, where several 
additional complications occur. Shifting the contour to the imaginary 
axis we get the pieces 
\beq
\zeta_{\delta} ^{p} (f,s) &=&\frac {\sin \pi s }{\pi}
\sumt\sum_{m=0}^{p-1} \sum_{l=0}^m  \delta_{ml}^{p}(-1)^m\, \nu^{-2s-2l}
\label{IV10}\\
& &\int_0^{\infty}
dz \,\, \frac{z^{-2s-2m}}{(\ug)^2 -\nu^2(1+z^2)} 
\frac{\partial}{\partial z} \ln \big((\ug )I_{\nu} (z\nu) +z\nu
I_{\nu} ' (\nu z) \big) \nn\\
\zeta_{\rm shift}^p (f,s) &=& -\frac {1}{ 2} \sum_{m=0}^{p-1}
\sum_{l=0}^m \delta_{ml}^p \sumt
\nu^{2m-2l} (\nu ^2 - (\ug)^2)
^{-s-m-1/2} \label{IV11}\\
& &\quad\quad \frac{\partial}{\partial k} \ln\big(kJ_{\nu}' (k) +
(\ug )J_{\nu} (k)\big) |_{k=\sqrt{\nu^2 -(\ug )^2} }, \nn
\eeq
the last one arising on moving the contour over the pole at 
$k=\sqrt{\nu^2-(\ug )^2}$.

In dealing with $\ze_\de^p (f,s)$ one can use, as done after eq. (\ref{III9}),
the uniform asymptotics of the Bessel functions and define analogously to 
(\ref{III11}) the asymptotic contributions $A_{i,\de} (f,s)$. We will 
illustrate the calculation by dealing with 
\beq
A_{-1,\de}^p (f,s) &=& \frac{\sin \pi s}{\pi} \sumt \smpe (-1)^m 
 \nu^{-2s-2l+1} \nn\\
& &\int_0^{\infty} dz \,\,  \frac{z^{-2s-2m-1}}{(\ug)^2 -\nu^2(1+z^2)} 
(1+z^2)^{1/2} .\nn
\eeq
By using the expansion
\beq
\frac 1 {(\ug)^2 -\nu^2 (1+z^2)} = -\sum_{i=0}^{\infty} \frac {(\ug )^{2i}}
{\nu^{2i+2} (1+z^2)^{i+1}} \nn
\eeq
the above integrals are recognised as representations of Beta functions 
\cite{45}.
As an intermediate result one gets
\beq
A_{-1,\de}^p (f,s) &=& \sum_{i=0}^{\infty}  \smpe 
\frac{\Gamma (s+i+m+1/2 )}{\Gamma (s+m+1) \Gamma (i+1/2)} \nn\\
& & \sumt \frac {(\ug )^{2i}} {\nu^{2s+2l+2i+1}} .\label{IV12}
\eeq
Complicated as the expression (\ref{IV12}) is, we have to remind ourselves of
our initial goal, namely, the determination of the heat-kernel coefficients 
up to $a_{3/2}$. Thus we need to determine only the residues of (\ref{IV12}) 
at $s=D/2,(D-1)/2,D/2-1$ and $(D-3)/2$. 

In general, the numbers $\de_{ml}^p$ contain terms independent of $n_d$ and 
linear in $n_d$, 
\beq
\de_{ml}^p = \de_{ml0}^p +\de_{ml1}^p g n_d .\nn
\eeq
Furthermore it is clear that the higher the power of $n_d$ the more to 
the right the pole of the associated term will be. Thus in addition consider
the expansion
\beq
(\ug )^{2i} = g^{2i} n_d^{2i} +2iu g^{2i-1} n_d ^{2i-1} + \cao (n_d ^{2i-2}) \nn
\eeq
in powers of $n_d$. With eq. (\ref{III23}) it is then obvious that the 
rightmost pole in $A_{-1,\de}^p (f,s) $ due to the 
$\cao (n_d ^{2i-2})$ term is situated at $s=(D-4)/2$ and 
contributes only to $A_2$. For our immediate purposes it is thus sufficient
to take into consideration only the above two terms. As a result
\beq
A_{-1,\de}^p (f,s) = F_1^p (f,s) +F_2 ^p (f,s) \nn
\eeq
with 
\beq
F_1^p (f,s) &=& \frac 1 2 \sum_{i=0}^{\infty} \smpen  
\frac{\Gamma (s+i+m+1/2)} {\Gamma (s+m+1) \Gamma (i+1/2)} \nn\\
& & g^{2i} E_{2i} (s+l+i+1/2) , \label{IV13}\\
F_2^p (f,s) &=& \frac 1 2 \sum_{i=0}^{\infty} \smpee 
 2iu \frac{\Gamma (s+i+m+1/2)} {\Gamma (s+m+1) \Gamma (i+1/2)} \nn\\
& & g^{2i} E_{2i} (s+l+i+1/2).  \label{IV14}
\eeq
Use of the residues of the base zeta function $E_{2i}$ then easily gives  
the following normalized contributions (we use the notation $\de^p_{ml}=0$ 
for $l<0$) 
\beq
\lefteqn{
\hspace{-3cm}\Gamma (D/2 -k) \frac{(4\pi)^{D/2}}
{(2\pi )^d} \mbox{Res } F_1^p 
(f,D/2-k) = 
\sum_{m=k-1}^{p-1} 
 \de_{m(k-1)0}^p \frac{\left(\frac d 2 \right)_{m+1-k} }{ 
(D/2-k)_{m+1}}     }\nn\\
& &\hspace{-2cm}{_2F_1} (1,D/2-k+m+1/2,
d/2; g^2) \label{IV15}
\eeq
\beq
\lefteqn{
\hspace{-3cm}\Gamma (D/2 -k) \frac{(4\pi)^{D/2}}
{(2\pi )^d} \mbox{Res } F_2^p
(f,D/2-k) =  
u \sum_{m=k-1}^{p-1} 
\de_{m(k-1)1}^p 
\frac{(d/2)_{m+1-k} }{  
(D/2-k)_{m+1}}    }\nn\\
& &\hspace{-2cm}g \frac d {dg}  \,\,
{_2F_1} (1,D/2-k+m+1/2,
d/2; g^2) \label{IV16}
\eeq
For our purposes, only $k=1$ is relevant, but we have
given these general results to show that in principle one could go further.

The same kind of argument allows one to show that the relevant parts in the
other $A_i ^\de (f,s)$ can all be representated in terms of hypergeometric 
functions. 

Finally we are left with the treatment of $\ze_{shift}^p (f,s)$, 
eq. (\ref{IV11}). Here, instead of the uniform asymptotics of $I_\nu$ we need 
it for the function $J_\nu$. The expansion given in \cite{44} (see also 
\cite{38}) suggests for $\nu\to\infty$,
\beq
\frac{\pa}{\pa k} \ln (kJ_\nu ' (k) +(\ug )J_\nu (k) ) \left|_{k=\sqrt{
\nu^2-(\ug )^2 } } \sim 
\sum_{l=0}^\infty e_l \frac{( \ug )^{2l+1}} {\nu^{2l+1}}\right. ,\label{IV17}
\eeq
with the coefficients $e_l$ to be determined. Here the problem appears,
that {\it every} value of $l$ contributes to the pole of $\ze_{shift}^p
(f,s)$ already at $s=(D-1)/2$. Thus, the asymptotic expansion of the left 
hand side to any power in $n_d /\nu$ is needed, apparently an extremely 
difficult problem on asymptotics of special functions. However, given that 
$\ze_{shift}^p (f,s)$ contributes only for $p>0$ we will show in the 
next section how to circumvent a direct evaluation of eq. (\ref{IV17}).

\section{Results and conformal techniques}
After having shown in detail several aspects of the special case calculation
on the generalized cone let us collect the information about the universal 
constants appearing in eqs. (\ref{II6}) -- (\ref{II8}). 

In order to answer the open question about eq. (\ref{IV17}) let us 
consider first $a_{1/2}$ and take $f(r) = f_0 $. For this case only 
$A_0 (s)$ and $A_+ (s)$ contributes and the answer is 
\beq
\frac{(4\pi)^{d/2}}{(2\pi )^d}a_{1/2} (f) = f_0 \frac 1 4 
\left( \frac 2 {\sqrt{1-g^2}} -1 \right) . \label{V1}
\eeq
Comparison with (\ref{II6}) gives the correct universal constant  
\beq
\de = \frac 1 4 \left( \frac 2 {\sqrt{1+\G^2}} -1 \right) \label{V2}
\eeq
and the present special case evaluation determines the $a_{1/2}$ 
coefficient for a general manifold (which was clear of course). Thus, for 
the case of the cone, taking $f(r) = f_0 +f_1 r^2$ we know that 
\beq
\frac{(4\pi)^{d/2}}{(2\pi )^d}a_{1/2} (f) = (f_0 +f_1 )\frac 1 4
\left( \frac 2 {\sqrt{1-g^2}} -1 \right) .\nn
\eeq
Taking into account all terms but the unknown contribution from 
$\ze_{shift}^p (f,s)$ we find 
\beq
(f_0 +f_1 ) \frac 1 4
\left( \frac 2 {\sqrt{1-g^2}} -1 \right) +\frac{g^2}{3d(1-g^2)^{3/2}} f_1. \nn
\eeq
Since the first piece is the correct answer, the last piece {\it has} to be
cancelled by the contribution of $\ze_{shift}^p (f,s)$. Use of the expansion 
(\ref{IV17}) and (\ref{IV11}) gives 
\beq
\G \left(\frac{D-1} 2 \right) \frac{(4\pi)^{d/2}}{(2\pi )^d} 
\mbox{Res } \ze_{shift}^p (f,(D-1)/2) = -\frac{g^2} {3d}f_1 
\sum_{i=0}^\infty \sum_{l=0}^\infty \frac{\left( \frac 3 2 \right) _{i+l} 
\left(\frac{d+1} 2 \right)_i }{\left(\frac d 2 +1\right)_{i+l} i!} 
e_l g^{2i+2l} 
\nn
\eeq
which leads to the condition 
\beq
\sum_{i=0}^\infty \sum_{l=0}^\infty \frac{\left( \frac 3 2 \right) _{i+l}
\left(\frac{d+1} 2 \right)_i }{\left(\frac d 2 +1\right)_{i+l} i!}
e_l g^{2i+2l}
=\frac 1 {(1-g^2)^{3/2}} .\nn
\eeq
In order that the left hand side should be nothing other than a complicated 
series 
expansion of the right hand side one has to conclude that $e_l = (1/2)_l /l! 
$ and one finds the surprisingly easy expansion
\beq
\frac{\pa}{\pa k} \ln (kJ_\nu ' (k) +(\ug ) ) \left|_{k=\sqrt{\nu^2 -(\ug )^2}}
\right. = \sum_{l=0}^\infty \frac{(1/2)_l}{l!} \frac{ (\ug )^{2l+1}}
{\nu^{2l+1}} . \label{V3}
\eeq
But having expansion (\ref{V3}) at hand the contribution of $\ze _{shift} ^p 
(f,s)$ to any pole can be determined so that now for the cone
complete knowledge for $a_1$ and $a_{3/2}$ is available. 

We turn now to $a_1$. Dealing first with $f(r) = f_0$, the 
linear term in $d$ defines $b_0$, the linear term in $S$ defines $b_2$,
the term independent of $d$ and $S$ defines $\sigma_1$. Dealing afterwards with
$f(r) = f_1 r^2$ the additional piece is immediately identified with $b_1$. 
As a result we obtain the correct answer 
\beq
b_0 &=& 2 - 6\,\left( -{\frac{1}{1 + {\G^2}}} + 
      {\frac{\mbox{Arctanh}({\sqrt{-{\G^2}}})}
        {{\sqrt{-{{{\G}}^2}}}}} \right),\nn \\
b_1 &=&  3 - {\frac{6\,\mbox{ArcTanh}({\sqrt{-{\G^2}}})}
{{\sqrt{-{\G^2}}}}},\nn \\
b_2 &=& {\frac{12}{1 + {\G^2}}},\nn  \\
\sigma_1 &=&  \frac{6}{\G^2}\left( -{\frac{1}{1 + {\G^2}}} + 
        {\frac{\mbox{ArcTanh}({\sqrt{-{\G^2}}})}
          {{\sqrt{-{{{\G}}^2}}}}} \right),\nn
\eeq
which shows that the ideas involved in our special case calculation are indeed
correct. Especially the example once more determines  the complete 
coefficient for a general smooth manifold with boundary.

Proceeding in the same way as described for $a_1$, we obtain the 
following universal constants for $a_{3/2}$, 
\beq
c_5&=&\frac 1 {\G^4} \left[
       2\,\left( -\left( {\G^2}\,
           \left( 144 - {\frac{160}{{\sqrt{1 + {\G^2}}}}} \right)  \right)  + 
        32\,\left( -1 + {\frac{1}{{\sqrt{1 + {\G^2}}}}} \right) \right.
            \right. \nn\\
& & \left. \left.         + 
        {\G^4}\,\left( -15 + {\frac{80}{{\sqrt{1 + {\G^2}}}}} \right)  \right) 
      \right],\label{V5}\\
c_6 &=&\frac 1 {\G^4} \left[ 
8\,\left( 32 - {\frac{32}{{\sqrt{1 + {\G^2}}}}} + 
         {\G^4}\,\left( -3 - {\frac{8}{{\sqrt{1 + {\G^2}}}}} \right)  - 
         {\G^2}\,\left( -36 + {\frac{52}{{\sqrt{1 + {\G^2}}}}} \right)  \right) 
              \right]\nn\\
 & &        + {\frac{32\,\left( 5\,{\G^4} - 
         8\,\left( -1 + {\sqrt{1 + {\G^2}}} \right)  - 
         4\,{\G^2}\,\left( -4 + 3\,{\sqrt{1 + {\G^2}}} \right)  \right) }{
       {\G^4}\,{\sqrt{1 + {\G^2}}}}} ,\label{V6} \\
c_7 &=& {\frac{192\,\left( 1 - {\sqrt{1 + {\G^2}}} - 
        {\G^2}\,\left( -2 + {\sqrt{1 + {\G^2}}} \right)  \right) }{{\G^2}\,
      {{\left( 1 + {\G^2} \right) }^{{\frac{3}{2}}}}}}, \label{V7}\\
c_8 &=& {\frac{192}{{{\left( 1 + {\G^2} \right) }^{{\frac{3}{2}}}}}} ,
\label{V8} \\
c_9 &=&  \frac{-192}{\G^2}\left(1 - {\frac{1}
             {{\sqrt{1 + {\G^2}}}}} \right), \label{V9}  \\
\beta_1 &=&  {\frac{-32\,\left( 5\,{\G^4} - 8\,\left( -1 + 
{\sqrt{1 + {\G^2}}} \right)  - 
        4\,{\G^2}\,\left( -4 + 3\,{\sqrt{1 + {\G^2}}} \right) 
         \right) }{{\G^6}\,
      {\sqrt{1 + {\G^2}}}}}, \label{V10}\\
\lambda_1 &=&  {\frac{192\,\left( -\left( {\G^2}\,
           \left( 3 - 2\,{\sqrt{1 + {\G^2}}} \right)  \right)  + 
        2\,\left( -1 + {\sqrt{1 + {\G^2}}} \right)  \right) }{{\G^4}\,
      {{\left( 1 + {\G^2} \right) }^{{\frac{3}{2}}}}}},\label{V11} 
\eeq
respectively the following relations among them,
\beq 
\lefteqn{
\hspace{-2cm}c_3-c_1+c_0/4= \frac 1 {\G^4 (1+\G^2)^{3/2}}  \left[
{\G^2}\,\left( 240 - 224\,{\sqrt{1 + {\G^2}}} \right)  +
      {\G^4}\,\left( 336 - 207\,{\sqrt{1 + {\G^2}}} \right)\right. } \nn\\ 
& &\left.      -
      32\,\left( -1 + {\sqrt{1 + {\G^2}}} \right)  -
      5\,{\G^6}\,\left( -16 + 3\,{\sqrt{1 + {\G^2}}} \right)   
\right] ,
        \label{V4}
\eeq
\beq
\lefteqn{
\hspace{-2cm}\G^2 (\sigma_3 -\mu_2 ) +c_4+c_1-c_0/2 = 
          \frac 6 {\G^4 (1+\G^2)^{3/2}} \left[
 32\,\left( -1 + {\sqrt{1 + {\G^2}}} \right) \right. } \nn\\
     & &    +
        {\G^6}\,\left( -48 + 7\,{\sqrt{1 + {\G^2}}} \right)  +
        16\,{\G^2}\,\left( -10 + 9\,{\sqrt{1 + {\G^2}}} \right) \nn\\
    & &\left.  +
        {\G^4}\,\left( -192 + 119\,{\sqrt{1 + {\G^2}}} \right)    
\right] ,
          \label{V12}
\eeq
\beq
\lefteqn{
\hspace{-2cm}\sigma_2 +\frac 1 {\G^2} (\sigma_4 +\mu_2 )
           +\frac {c_0}{4\G^4} = -\frac 8 {\G^8 
        (1+\G^2)^{3/2} } \left[ 
 32\,\left( -1 + {\sqrt{1 + {\G^2}}} \right)\right. } \nn\\
   & & +
        {\G^6}\,\left( -32 + 3\,{\sqrt{1 + {\G^2}}} \right)  +
        8\,{\G^2}\,\left( -15 + 13\,{\sqrt{1 + {\G^2}}} \right)\nn\\
   & &\left.  +
        3\,{\G^4}\,\left( -42 + 25\,{\sqrt{1 + {\G^2}}} \right)      
\right] . 
      \label{V13}
\eeq
Where applicable, in the limit $\G \to 0$ the results for Robin boundary 
conditions are reproduced as a check.

This serves as a very good input for applying the techniques of \cite{12}.
First we use a result on product manifolds \cite{12}, which in our 
case gives
\beq
c_0 &=& 96\,\left( -1 + {\frac{2}{{\sqrt{1 + {\G^2}}}}} \right), \label{V14}\\
c_1 &=&  16\,\left( -1 + {\frac{2}{{\sqrt{1 + {\G^2}}}}} \right), \label{V15}
\eeq
Together with eq. (\ref{V4}) this also determines $c_3$,
\beq
c_3&=& \frac 1 {\G^4 (1+\G^2)^{3/2}} 
     \left[\G^2\,\left( 240 - 224\,{\sqrt{1 + {\G^2}}} \right)  + 
      {\G^4}\,\left( 320 - 199\,{\sqrt{1 + {\G^2}}} \right) \right.\nn\\
  & &\left. + 
      {\G^6}\,\left( 64 - 7\,{\sqrt{1 + {\G^2}}} \right)  - 
      32\,\left( -1 + {\sqrt{1 + {\G^2}}} \right)  \right].
\label{c3}
\eeq
The remaining task is to apply the functorial techniques of 
\cite{12}. For details of the technique itself see this reference and for 
the modifications when tangential derivatives are involved see \cite{31}.

The basic equations are the conformal-variation formulae 
\beq
\frac d {d\ep} \left|_{\ep =0}\right.  a_{n/2} (1,e^{-2\ep f}L ) &=& (D-n) 
                               a_{n/2} (f,L ) ,\label{V16} \\
\frac d {d\ep} \left|_{\ep =0}\right. a_{n/2} (e^{-2\ep f}H, e^{-2\ep f} L) 
         &=& 0 \mbox{ for }D=n+2, \label{V17}
\eeq
with an arbitrary smooth function $H$. For a collection of variational 
formulae again see \cite{12}. The additional relation $\Gamma^i (\ep) = 
e^{-\ep f}\Gamma^i$ is given in \cite{31}. Setting to zero the 
coefficients of all terms in (\ref{V16}) for $n=3$ one obtains, for example 
\cite{31},
\beq
\begin{array}{ll}
\underline{\mbox{Term}} & \underline{\mbox{Coefficient}} \\
f_{;NN} & 0=\frac 1 2 (D-2)c_0 -2(D-1)c_1 -(D-1)c_2 -(D-3)c_6 -\G^2\mu_1 \\
Kf_{;N} & 0=\frac 1 2 (D-2)c_0 -2(D-1)c_1 -c_2+2(D-1)c_3 +2c_4 \\
& \phantom{0=}-\frac 1 2 (D-2)c_7 -(D-3)c_5 +\G^2\sigma_3 -\G^2 \mu_2 
\end{array}\nn
\eeq
The first of these determines $c_2$ and $\mu_1$, namely 
\beq
c_2&=&\frac{8}{\G^2}\left( 12 - {\frac{12}{{\sqrt{1 + {\G^2}}}}} + 
        {\G^2}\,\left( 1 - {\frac{8}{{\sqrt{1 + {\G^2}}}}} \right)  \right) 
    ,  \label{V18}\\
\mu_1 &=& {\frac{96\,\left( 2 + {\G^2} - 2\,{\sqrt{1 + {\G^2}}} \right) }
    {{\G^4}\,{\sqrt{1 + {\G^2}}}}}. \label{V19}
\eeq
The second together with (\ref{V5}) gives 
\beq
c_4 = \frac 2 {\G^4}   
\left( {\G^4}\,\left( 5 - {\frac{32}
           {{\sqrt{1 + {\G^2}}}}} \right)
            + {\G^2}\,\left( 48 - {\frac{32}{{\sqrt{1 + {\G^2}}}}} \right)  + 
        32\,\left( -1 + {\frac{1}{{\sqrt{1 + {\G^2}}}}} \right)  \right) 
     \label{V20}
\eeq
In addition we get 
\beq
b_1 = 0\label{neu}.
\eeq
Disappointing as it is, under the given assumptions these are the only 
new universal constants the functorial techniques yield. But {\it due}
to the restrictions imposed we have not yet exploited all information 
available. For example one has the variational formula
\beq
\frac d {d\ep} \left|_{\ep =0} \right. 
        R^l_{\phantom{l}ilj} \G^i\G^j &=& -2fR^l_{\phantom{l}ilj} \G^i\G^j
                    -(D-3) K_{ij} \G^i\G^j f_{;N} \label{V21}\\
       & &-\G^2 K f_{;N} -(D-3) f_{:ij} \G^i\G^j -\G^2 f_{:ll}. \nn
\eeq
Up to now we have not compared coefficients involving tangential 
derivatives, because for covariantly constant $\G^i$ these pieces 
integrate to zero. But if we relax the condition of covariantly constant 
$\G^i$, eq. (\ref{V21}) shows that two additional eqs. for the universal 
constant $\mu_2$ would arise. (In contrast the variations of the terms
associated with the missing $\sigma_2,\sigma_3$ and $\sigma_4$ do not 
contain any tangential derivatives.) In order to exploit this observation
we have to generalize eq. (\ref{II4}) to when  
$\widehat\nabla_j\G^i \neq 0$. This is done as usual by building up all 
possible independent geometrical terms with certain homogeneity properties
\cite{12}. For the boundary conditions under consideration eq. (\ref{II4}) 
has to be supplemented by the following terms,
\beq
\frac{384}{(4\pi )^{1/2}}a_{3/2}^{cov} (f) &=& 
                  Tr[f 
               ( \gamma_1 \G^i_{\phantom{i}:j} \G_{i:}^{\phantom{j}}+
                  \gamma_2 \G^i_{\phantom{i}:j} \G^j_{\ph{j}:i}+
                  \ga_3 \G^i_{\ph{i}:i}\G^j_{\ph{j}:j} \nn\\
             & &+\ga_4\G^i_{\ph{i}:ij}\G^j +\ga_5 \G_{i:j}^{\ph{i:j}j}\G^i+
                \ga_6 \G^{i}_{\ph{i}:j}\G_{i:k} \G^j\G^k\nn\\
        & &+\ga_7 \G^i_{\ph{i}:j} \G_{k:i} \G^j\G^k +
              \ga_8 \G^i_{\ph{i}:j}\G_{k:}^{\ph{k:}j}\G_i \G^k +
             \ga_9 \G^i_{\ph{i}:i} \G^l_{\ph{l}:k}\G_l \G^k \nn\\
         & &\ga_{10} \G^i_{\ph{i}:jk} \G_i \G^j\G^k +\ga_{11}
              \G^i_{\ph{i}:j}\G^l_{\ph{l}:k}\G^j\G^k\G_i\G_l )] [\pa \cam].
                   \label{V22}
\eeq
The term $\G^i_{\ph{i}:ji}\G^j$ is not added because due to the Gauss-Codacci
relation one has 
\beq
\G^i_{\ph{i}:ji} \G^j = \G^i_{\ph{i}:ij}\G^j +R^i_{\ph{i}kij}\G^k\G^j +
                  K K_{kj}\G^k\G^j -K_{ki} K^i_j \G^k\G^j \nn
\eeq
so that this term is linearly dependent on the others already displayed.
Because of the simple conformal transformation property of $\G^i$ it is 
relatively easy to find the variational formulas of all invariants in 
eq. (\ref{V22}). Even though we know 
none of the $\ga_i$, setting to zero the coefficients of the tangential 
derivatives terms in (\ref{V21}), we find the unambigous answer 
\beq
\mu_2 = 0 .\label{V23} 
\eeq
As a consequence eq. (\ref{V5}) shows 
\beq
\sigma_3 &=& \frac 1 {\G^6\left( 1 + {\G^2} \right)^{3/2}}
       \left[32\,\left( -5\,{\G^6} + 8\,
           \left( -1 + {\sqrt{1 + {\G^2}}} \right)  + 
        6\,{\G^4}\,\left( -5 + 3\,{\sqrt{1 + {\G^2}}} \right)
          \right.\right. \nn\\
      & &\left.\left.  + 
        {\G^2}\,\left( -30 + 26\,{\sqrt{1 + {\G^2}}} \right)  
              \right) \right] 
       \label{V24}
\eeq

This is really all we can get from the lemmas and the 
specific example of the generalized cone, $I\times T^d$. 
We can obtain information only about the 
combination $\G^2 \sigma_2 +\sigma_4$. Thus some additional input is needed
to accomplish the goal of finding all universal constants in (\ref{II4}).

\section{Oblique boundary conditions on $B^2\times T^{d-1}$}
One possibility of finding the remaining information is to look for an 
example which is able to separate the contributions of 
$(K_{ij}\G^i \G^j)^2$ 
and $K_{ij} K^j_l \G^i\G^l$. The reason that all types of generalized 
cones with metric (\ref{II5}) fail to do so, is that $K_i^j = \de_i^j$. As 
a result,  the contraction $K_{ij} \G^i\G^j$ equals  $\G^2$ and also 
$K_{ij} K^j_l \G^i\G^l = \G^2$. Having a term like $\sigma_2 (\G^2) 
\G^4 +\sigma_4 (\G^2) \G^2 = g(\G^2)$, with $g(\G^2) $ a known function of 
$\G^2$, there is no possibility  uniquely determining $\sigma_2$ or 
$\sigma_4$ because, as indicated, these also
depend on $\G^2$. It is clear that this 
problem is not a result of having chosen only {\it one} non-vanishing 
component $\G^i$. For a generalized cone, these properties are generic. 
So we are forced to leave this class of examples.

The difference between the invariants 
associated with $\sigma_2$ and $\sigma_4$ is that the first contains 
fourth powers of $\G^i$ whereas the second one only squares. If we deal 
with two instead of one nonvanishing component, say $\G^d = g$ and 
$\G^i = g_d$, $i\neq d$, $g_d,g$ constants, and if the extrinsic curvature 
could be a projector on one of them, say $K_{dd}=1$, $K_{ij}=0$ for 
$(i,j)\neq (d,d)$, then $(K_{ij}\G^i\G^j)^2 = g_d^4$ and 
$K_{ij}K^j_l \G^i\G^l = g_d^2$. By simply comparing powers of $g_d$ 
the universal constant $\sigma_2$ could be determined being the only 
one with $g_d^4$. 

Using the metric 
\beq
ds^2 = dr^2 +d\Sigma^2 \nn
\eeq
and $K_{ab} = -\Gamma^r_{ab}$, $\Gamma^r_{ab}$ being the Christoffel symbol,
it is seen that keeping the manifold topologically as $I\times T^d$, 
the metric 
\beq
d\Sigma^2 = dx_1^2 +...+dx_{d-1}^2 +r^2 dx_d^2 \nn
\eeq
will have the above property. This is clearly the flat manifold 
$B^2\times T^{d-1}$ and the eigenvalue problem is easily solved. With the 
notation $\nt = n_1^2+...+n_{d-1}^2$ the eigenfunctions are
\beq
J_{|n_d|} (r\sqrt{\alpha^2 -\nt }) e^{i(x_1n_1 + ... + x_d n_d )} ,\,\,
                    \nt \in \intgs ^d \nn
\eeq
with eigenvalues $\alpha^2$. The boundary condition takes the form
\beq
\sqrt{\alpha^2 -\nt }J_{|n_d|} '(\sqrt{\alpha^2 -\nt }) + 
(g_dn_d +gn-S) J_{|n_d|}(\sqrt{\alpha^2 -\nt }) =0 ,\label{VI1}
\eeq
where we have used $n =n_i$ (the result is the same for any 
$i\in \{1,...,d-1\}$).

Our main interest is to determine $\sigma_2$ and $\sigma_4$. The calculation 
involving two tangential derivatives will be seen to be sufficiently 
difficult so that we will restrict ourselves to what is strictly necessary, 
namely we will not
bother to do the smeared calculation but will be content with the 
sufficient choice $f(r) = 1$. The information derived about
$c_3+c_4,c_7,c_8,\lambda_1$ will serve as a further check of the 
previous calculation, the new quantities, $\sigma_2$ and $\sigma_3+\sigma_4$ 
will complete our analysis of $a_{3/2} (f) $, for covariantly constant 
$\G^i$. 

One can procede very much as before. Starting with a contour representation 
similar to eq. (\ref{III8}) and shifting the contour to the imaginary axis, 
one gets 
\beq
\ze (s) =\frac{\sin \pi s} \pi \sumt \int_{|\vec n_t|}^\infty 
 dk \,\, (k^2 -\nt )^{-s} \frac{\pa}{\pa k} \ln 
\left( kI_{|n_d|}' (k) + [g_dn_d +gn -S] I_{|n_d|} (k) \right).\label{VI2}
\eeq
It is seen, that $\nt$ acts effectively as a mass of the field and comparing
with the previously treated example $n_d$ plays the role of $\nu$.

The general procedure of dealing with $\ze (s)$ is the same as in 
section 3. However several complications arise, and the situation is 
sufficiently different so as to warrant further describtion.

One starts from the uniform asymptotic expansion of the Bessel function 
\cite{45} and eq. (\ref{III10}) is found with the replacements 
$gn_d\to g_dn_d +gn$, 
$u\to -S$ and $\nu \to |n_d|$. Asymptotically one again finds 
eq. (\ref{III11}) with the characteristic differences already described. 
The several new features arising are now dealt with by looking at 
\beq
A_+ (s) &=& \frac{\sin \pi s}{\pi} \sumtn\,\, \sumnd  \quad
             \int_{|\vec n_t/n_d|}^
     \infty dz \,\, [z^2n_d^2 -\nt]^{-s} \label{VI3}\\
  & & \frac \pa {\pa z} \ln\left( 1 +\frac {gn+g_dn_d}{|n_d|} 
\frac 1 {\sqrt{1+z^2}} \right).\nn
\eeq
The integral is nothing but a hypergeometric function \cite{45} and we get
\beq
A_+ (s) &=& -\frac 1 {2\G (s)} \sum_{l=0}^\infty (-1)^l \frac{\G (s+(l+1)/2)}
        {\G ((l+3)/2) } \nn\\
   & &\sumtn \,\, \sumnd \quad (gn+g_dn_d)^{l+1} 
               (\nt )^{-s-(l+1)/2}  \nn\\
   & &\hspace{2cm}      {_2F_1} \left( 
         \frac{l+3} 2 , s+\frac{l+1} 2, \frac{l+3} 2; -\left| \frac 
          {n_d}{\vec n_t} \right| ^2 \right). \label{VI4}
\eeq
The apparent difficulty is to extract the meromorphic structure of multiple 
sums of hypergeometric functions. This is very effectively done by using 
the Mellin-Barnes integral representation of $_2F_1$ \cite{45},
\beq
_2F_1 (\alpha , \beta , \gamma; z) =\frac{\G (\ga )}{\G (\alpha ) \G (\beta)}
\frac 1 {2\pi i} \int_{-i\infty}^{i\infty} dt\,\,
     \frac{\G (\alpha +t) \G (\beta +t) \G (-t)}{\G (\ga +t)} (-z)^t,\nn
\eeq
where the contour is chosen such that the poles of the function $\G (\alpha 
+t)$ and $\G (\beta +t)$ lie to the left of the path of integration and the 
poles of the function $\G (-t)$ lie to the right of it. When using this 
integral representation it is seen that the sum over $\vec n_t$ leads to
$(d-1)$-dimensional Epstein type zeta functions, whereas the sum over $n_d$ 
gives a Riemann zeta function. The relevant 
zeta function is again of the Epstein type (\ref{III7}) which we now write
as
\beq
E_{t,2l} (s) = \sumtn (\nt )^{-s} n^{2l} .\label{VI5}
\eeq
As a result of the described steps one arrives at 
\beq
A_+ (s) &=& \frac 1 {\G (s)} \sum_{l=1}^\infty \sum_{k=0}^l \frac 1 {l!}
   \left(
          \begin{array}{ll}
              2l \\
              2k 
          \end{array}  \right)  g^{2k} g_d ^{2l-2k} \label{VI6} \\
   & &\frac 1 {2\pi i} \int_{-i\infty}^{i\infty} dt\,\,
         \G (s+l+t) \G (-t) \ze_R (-2t-2l+2k) E_{t,2k} (s+l+t),\nn
\eeq
where the contour (depending on $l$ and $k$) is such that the poles of 
$\ze_R$ lie to the right of the contour, the poles of $E_{t,2k}$ to the 
left
of it. This Mellin-Barnes representation allows the 
meromorphic structure of $A_+ (s)$ to be read off by closing the contour 
to the left. 
We then encounter poles of $\G (s+l+t)$ at $t=-s-l-m$, $m\in \nats_0$ with
residues $\G (s+l+m) \ze_R (2s+2k+2m) E_{t,2k} (m)$. The right-most pole 
lies at $s=1/2$ and it is clear that the poles of the $\G$-function are 
irrelevant for our purposes. However, the pole of the Epstein function is 
situated at $t=(d-1)/2 -l-s+k$ and gives relevant contributions.

Keeping only these terms,
\beq
A_+ (s)& \sim & \frac{\pi^{(d-1)/2}}{\G (s)} \sum_{l=1}^\infty \frac 1 {l!}
            \sum_{k=0}^l \left(
          \begin{array}{ll}
              2l \\
              2k
          \end{array}  \right) \frac {\G (k+1/2)} {\sqrt{\pi}} 
         \G (s+l-k-(d-1)/2 ) \nn\\
    & &\hspace{2cm}\ze_R (2s-d+1) g^{2k} g_d ^{2l-2k}.\label{VI7}
\eeq
The right most pole at $s=d/2$ comes from the Riemann zeta function. Using 
the relations \cite{45}
\beq
\frac{\G (x)}{\G (2x)} = \frac{\sqrt{\pi}}{2^{2x-1}\G (x+1/2)}\nn
\eeq
and 
\beq
\frac{(2l)!}{l! 2^{2l}} = \frac{(2l-1)!}{2^l} = \frac{\G (l+1/2) }
        {\sqrt{\pi}} \nn
\eeq
one gets
\beq
\mbox{Res }A_+ (d/2) = \frac{\pi^{d/2}}{2\G (d/2)} \left\{ (1-g^2 -g_d^2)
          ^{-1/2} -1\right\}  .\label{VI8}
\eeq
The calculation shows the manner in which $\G^2 = g^2 +g_d^2$ is built up.
Together with the contribution of $A_0 (s)$ one finds the correct 
coefficient $a_{1/2}$.

The next pole in (\ref{V17}) at $s=(d-1)/2$ comes from the $\G$-function 
for $k=l$. Then 
\beq
\mbox{Res }A_+ ( (d-1)/2) = -\frac{\pi^{(d-1)/2}}{2\G ((d-1)/2)} 
             \left\{ (1-g^2)^{-1/2} -1\right\}  .\label{VI9}
\eeq
This piece is cancelled by another contribution, but the example 
shows the way other contributions than $\G^2$ appear. 

There are no further (interesting) poles due to the zeroes of $\ze_R (s)$ 
at $s=-2m$, $m\in\nats$.

The basic characteristics, namely that the integrals are 
hypergeometric functions and that the eigenvalue sums may be dealt with
by Mellin-Barnes integral representations of these, are present for 
all other $A_j (s)$. The final results can all be written in terms of 
$_2F_1$ or $_3F_2$ or their derivatives, very much as in section 3. 
For the sake of space it is impossible to give further details.

In summary we confirm our results for the universal constants 
$c_3+c_4, c_7,c_8
$ and $\lambda_1$. Most important, we determine the remaining constants 
to be
\beq
\sigma_2 &=& \frac 1 {\G^8  \left( 1 + {\G^2} \right) ^{\frac{3}{2} } }
        \left[ -48\,\left( -5\,{\G^6} + 
        16\,\left( -1 + {\sqrt{1 + {\G^2}}} \right)  + 
        8\,{\G^2}\,\left( -5 + 4\,{\sqrt{1 + {\G^2}}} \right)
             \right.\right. \nn\\
         & & \left.\left.+ 
        {\G^4}\,\left( -30 + 16\,{\sqrt{1 + {\G^2}}} \right)  \right) \right]
        \label{VIII10}\\
\sigma_4 &=&   {\frac{32\,\left( -{\G^4} + 16\,
            \left( -1 + {\sqrt{1 + {\G^2}}} \right)  + 
        2\,{\G^2}\,\left( -7 + 3\,{\sqrt{1 + {\G^2}}} 
           \right)  \right) }{{\G^6}\,
      {\sqrt{1 + {\G^2}}}}}   \label{VIII11}
\eeq
and have thereby achieved the goal of determining $a_{3/2} (f)$ 
in eq. (\ref{II4}).
\section{Conclusions}
In this article we have developed a technique for the calculation of smeared
heat-kernel coefficients on generalized cones or manifolds of the type 
$B^n\times T^{D-n}$ for operators of Laplace type with oblique boundary 
conditions. These boundary conditions arise in response to questions in 
quantum gravity, gauge theory and string theory \cite{21,23,24,26,27,31}.
The asymptotic properties of the generalized boundary conditions encoded in
the asymptotic heat-kernel expansions are considerably more involved than
the corresponding ones for the traditional conditions. This 
article makes an attempt to provide a practical approach for the 
calculation of coefficients to any order needed although we suspect that
the involved analysis, complicated results and restrictions indicate that
further analysis along the lines of this paper should not be lightly
undertaken without strong motivation.

The direct calculation of higher coefficients for general curved manifolds
with arbitrary smooth boundaries becomes very difficult and impractical.
In the approach promoted here (see also \cite{38}) this analysis is 
avoided.
Based on the observation that functorial methods give relations 
between the numerical multipliers in the heat-kernel coefficients 
\cite{12} the remaining task is to find as many multipliers as needed 
by other means. A rich source of information is special case calculation.
Done in a systematic fashion in arbitrary dimensions and with a smearing 
function as general as necessary, we have seen that simple comparison 
with the general form of the coefficient yields the encoded information 
on the 
universal constants. Several checks on the constants found are 
provided by the conformal relations as well as by the different examples 
treated. A systematic feature of the approach is the use of algebraic 
computer programs.

It is clear that the approach can also be applied to the calculation of 
$a_2$, however, more than 100 terms are involved and a very real 
additional effort is necessary as well as a substantial motivation. 
Generalization to covariantly nonconstant
$\G^i$ is desirable and can be attacked by taking different base
manifolds (eg. a sphere) or by considering simple dependences of $\G^i$ 
on one of the tangential 
variables. Finally, it is hoped that also for the non-Abelian
setting new information can be obtained in the spirit of the 
present work.

\vspace{1cm}

{\bf Acknowledgments}

We wish to thank
Michael Bordag for very helpful discussions.  

This investigation has been partly supported by the DFG under contract 
number
BO1112/4-2.

%\begin{references}


\begin{thebibliography}{99}

\bibitem{1}
Gilkey P B,
{\em Invariance theory, the heat equation, and the Atiyah-Singer index
  theorem}.
2nd. Edn., CRC Press, Boca Raton 1995.

\bibitem{2}
Birrell N and Davies P C W,
{\em Quantum Fields in Curved Spaces}.
Cambridge,  (1982). Cambridge University Press.

\bibitem{3}
Buchbinder I L and Odintsov S D and Shapiro I L,
{\em Effective Action in Quantum Gravity}.
Bristol and Philadelphia,  (1992). IOP Publishing.

\bibitem{4}
Fulling S A, {\em Aspects of Quantum Field Theory in Curved 
Space-Time}, Cambridge (1989) Cambridge University Press.

\bibitem{5}
Moss I G, {\em Quantum Theory of Black Holes} (Ney York, Wiley, 1996).

\bibitem{6}
Esposito G, {\em Quantum Gravity, Quantum Cosmology and Lorentzian 
Geometries}. Lecture Notes in Physics, Monographs, Vol. m12, 
Berlin: Springer-Verlag, 1994.

\bibitem{7}
Greiner P,
Arch.~Rat.~Mech.~Anal., {\bf 41}, 163 (1971).

\bibitem{8}
Gilkey~P B,
J.~Diff.~Geom., {\bf 10}, 601 (1976).

\bibitem{9}
Avramidi~I G,
Nucl.~Phys.~B, {\bf 355}, 712 (1991).

\bibitem{10}
Fulling~S A and Kennedy G,
Transac.~Amer.~Math.~Soc., {\bf 310}, 583 (1988).

\bibitem{11}
Van den Ven A, {\em Index free Heat-kernel Coefficients},
hep-th/9708152.

\bibitem{12}
Branson~T P and Gilkey~P B,
Commun.~PDE, {\bf 15}, 245 (1990).

\bibitem{13}
Vassilevich D V,
J.~Math.~Phys. {\bf 36} 3174 (1995).

\bibitem{14}
Moss~I G and Dowker~J S,
Phys.~Lett.~B, {\bf 229}, 261 (1989).

\bibitem{15}
Mc~Avity~D M and Osborn H,
Class.~Quantum Grav., {\bf 8}, 603 (1991).

\bibitem{16}
Dettki A and Wipf A,
Nucl.~Phys.~B, {\bf 377}, 252 (1992).

\bibitem{17}
Dowker~J S and Schofield~J P,
J.~Math.~Phys., {\bf 31}, 808 (1990).

\bibitem{18}
Cognola G, Vanzo L, and Zerbini S,
Phys.~Lett.~B, {\bf 241}, 381 (1990).

\bibitem{19}
Branson T P, Gilkey P B and Vassilevich D V, 
Bolletino U.M.I. (7) {\bf 11B} Suppl. Fasc. 2 (1997) 39.

\bibitem{20}
Esposito G, Kamenshchik A Yu and Pollifrone G, {\em Euclidean Quantum 
Gravity on Manifolds with Boundary} (Fundamental Theories of Physics 85)
(Dordrecht: Kluwer, 1997).

\bibitem{21}
Barvinsky A O, Phys. Lett. B {\bf 195} (1987) 344.

\bibitem{22}
Marachevsky V N and Vassilevich D V, Class. Quantum Grav. {\bf 13} (1996) 645.

\bibitem{23}
Moss I G and Silva P J, Phys. Rev. D {\bf 55} (1997) 1072.

\bibitem{24}
McAvity D M and Osborn H, Class. Quantum Grav. {\bf 8} (1991) 1445.

\bibitem{25}
Avramidi I G, Esposito G and Kamenshchik A Yu, Class. Quantum Grav. {\bf 13} 
(1996) 2361.

\bibitem{26}
Abouelsaood A, Callan C G, Nappi C R and Yost S A, Nucl. Phys. B {\bf 280}
(1987) 599.

\bibitem{27}
Callan C G, Lovelace C, Nappi C R and Yost S A, Nucl. Phys. B {\bf 288} 
(1987) 525.

\bibitem{28}
Egorov Yu V and Shubin M A, {\em Partial Differential Equations} 
(Berlin: Springer, 1991).

\bibitem{29}
Krantz S G, {\em Partial Differential Equations and Complex Analysis} 
(Boca Raton, FL:CRC, 1992).

\bibitem{30}
Treves F, Introduction to Pseudodifferential and Fourier Integral Operators 
vol 1 (New York: Plenum, 1980). 

\bibitem{31}
Avramidi I G and Esposito G, Class. Quantum Grav. {\bf 15} (1998) 281.

\bibitem{32}
Bordag M, Kirsten K and Dowker S, Commun. Math. Phys. {\bf 182} (1996) 371.

\bibitem{33}
Bordag M, Elizalde E and Kirsten K, J. Math. Phys. {\bf 37} (1996) 895.

\bibitem{34}
Bordag M, Elizalde E, Geyer B and Kirsten K, Commun. Math. PHys. {\bf 179}
(1996) 215.

\bibitem{35}
Barvinsky A O, Kamenshchik Yu A and Karmazin I P, Ann. Phys. {\bf 219}
(1992) 201.

\bibitem{36}
Dowker J S, Class. Quantum Grav. {\bf 13} (1996) 585.

\bibitem{37}
Dowker J S, J. Math. Phys. {\bf 30} (1989) 770.

\bibitem{38}
Dowker J S and Kirsten K, Smeared heat-kernel coefficients on the ball 
and generalized cone, Preprint March 1998, hep-th/9803094.

\bibitem{39}
Dowker J S and Kirsten K, Class. Quantum Grav. {\bf 14} (1997) L169.

\bibitem{dimaeli}
E. Elizalde and D.V. Vassilevich, Chern-Simons boundary conditions 
in quantum electrodynamics, in preparation.

\bibitem{40}
Kirsten K, Class. Quantum Grav. {\bf 15} (1998) L5.

\bibitem{41}
Avramidi I G and Esposito G, Gauge theories on manifolds with boundary,
hep-th/9710048.

\bibitem{42}
Hawking S W and Ellis G F R, {\em The large scale structure of space time},
Cambridge University Press, Cambridge, 1973.

\bibitem{43}
Cheeger J, J. Diff. Geom. {\bf 18} (1983) 575.

\bibitem{44}
Abramowitz M and Stegun I A, {\em Handbook of Mathematical Functions},
New York, Dover, 1972.

\bibitem{45}
Gradshteyn I S and Ryzhik I M, {\em Tables of Integrals, Series and 
Products}, New York, Academic Press, 1965.

\bibitem{46}
Watson G N, {\em Theory of Bessel Functions}, Cambridge University Press, 
Cambridge, 1944.

%\end{references}
\end{thebibliography}
\end{document}